\documentclass[12pt]{iopart}

\usepackage{cite}
\usepackage{amssymb, amsfonts}
\usepackage{array}
\usepackage{multicol}
\usepackage{comment}
\usepackage{xcolor}
\usepackage{soul}
\usepackage{algorithmic}
\usepackage{graphicx}
\usepackage{textcomp}
\usepackage{tabularx}
\usepackage{caption}
\usepackage{subcaption}
\usepackage{hyperref}
\usepackage[USenglish]{babel}

\begin{document}

\title{Using EEG Signals to Assess Workload during Memory Retrieval in a Real-world Scenario}

\author{Kuan-Jung Chiang$^{\dag}$, Steven Dong$^{\ddag}$, Chung-Kuan Cheng$^{\dag}$, and Tzyy-Ping Jung$^{\S}$}
\ead{kuchiang@eng.ucsd.edu}
\address{$^{\dag}$ Department of Computer Science and Engineering, University of California San Diego}
\address{$^{\ddag}$ Human Factors Center of Excellence at Microsoft Corporation}
\address{$^{\S}$ Institute for Neural Computation and Institute of Engineering in Medicine, University of California San Diego}

\vspace{10pt}

\begin{abstract}
\textit{Objective}:
The Electroencephalogram (EEG) is gaining popularity as a physiological measure for neuroergonomics in human factor studies because it is objective, less prone to bias, and capable of assessing the dynamics of cognitive states. This study investigated the associations between memory workload and EEG during participants' typical office tasks on a single-monitor and dual-monitor arrangement. We expect a higher memory workload for the single-monitor arrangement.
\textit{Approach}:
 We designed an experiment that mimics the scenario of a subject performing some office work and examined whether the subjects experienced various levels of memory workload in two different office setups: 1) a single-monitor setup and 2) a dual-monitor setup. We used EEG band power, mutual information, and coherence as features to train machine learning models to classify high versus low memory workload states.
\textit{Main results}:
 The study results showed that these characteristics exhibited significant differences that were consistent across all participants. We also verified the robustness and consistency of these EEG signatures in a different data set collected during a Sternberg task in a prior study.
\textit{Significance}:
The study found the EEG correlates of memory workload across individuals, demonstrating the effectiveness of using EEG analysis in conducting real-world neuroergonomic studies.

\end{abstract}

\section{Introduction}
\label{sec:introduction}

The goal of human factors research is to build the best possible human-machine interface \cite{ms1992human}. The principles of human factors are applied to the design of systems to improve human comfort, productivity, and error reduction. Physical and cognitive ergonomics are two sub-fields of human factors research. Physical ergonomics is concerned with monitoring human peripherals such as muscle activity and tension. On the other hand, cognitive ergonomics, also known as neuroergonomics, assesses human cognitive states or brain activities, including memory and mental workload \cite{mehta2013neuroergonomics}.

Neuroergonomics has received increasing attention and has been combined with neuroscience research as brain-imaging methods have improved in recent years. To obtain human cognitive states, neuroergonomics has always relied on performance evaluation and subjective measurements \cite{wang2015using}. Performance metrics are indirect methods that infer a subject's cognitive workload based on how well they complete specific tasks. \cite{cegarra2008use}. Subjective measurements are commonly carried out by administering questionnaires to subjects.  As a result, these procedures are prone to bias. Furthermore, subjective measurements may only be obtained after specific activities have been finished, making tracking the dynamic of human cognitive states during tasks difficult. \cite{hart1988development}. Because of these constraints, human factor researchers have shifted their attention to physiological measures, which can provide continuous, direct, and objective measurements.

Several brain-imaging technologies, including functional magnetic resonance imaging (fMRI), functional near-infrared spectroscopy (fNIRS), and electroencephalography(EEG), are available as physiological measures to explore brain functions and neural activity. Both fMRI and fNIRS assess changes in blood oxygen in brain vessels and have lower temporal resolution \cite{sitaram2007fmri, naseer2015fnirs}. FMRI provides excellent spatial resolution and localization of activities of brain areas in a 3D space, however, due to the need for a strong magnet, it is almost exclusively used in laboratory settings  \cite{logothetis2008we}. The EEG monitors voltage fluctuations caused by the ionic current in the brain's neurons. Although EEG has a low spatial resolution, it has a high temporal resolution. EEG and fNIRS have several advantages, the most noteworthy of which is their portability. As a result, they have been employed in many studies to investigate ergonomics for a variety of tasks or settings \cite{aghajani2017measuring, bunce2011implementation, causse2017mental, herff2014mental, berka2007eeg, berka2005evaluation, brouwer2012estimating}.

Even though much research has been undertaken to investigate EEG or fNIRS correlates of memory or mental workload, only a few have focused on real-world scenarios. Participants in these studies completed tasks such as the n-back test, Sternberg task \cite{sternberg1966high}, or Simultaneous Capacity test \cite{bratfisch2008simkap}, which were meant to induce various levels of memory or mental effort. On the other hand, humans are unlikely to undertake such jobs in real life. As a result,  whether these neuroergonomic approaches can accurately represent the differences in cognitive states in a real-life scenario remains to be seen. Some studies looked into workload in the real world, although the tasks they used were mostly airplane control or human-robot interaction, which are still not typical daily tasks \cite{peck2014using, shou2013frontal}.
\color{black}
Furthermore, only a small percentage of them \cite{di2019eeg, giorgi2021wearable} are genuinely relevant to human factors research.
\color{black}

This study applied EEG analysis to a real-world scenario. More specifically, we designed an experiment to examine the memory workload when a subject performs office work under two different conditions: 1) a single-monitor setup and 2) a dual-monitor setup.
\color{black}
Because the dual-monitor setup can lessen the frequency of switching between windows, we think the memory usage can be decreased when performing specific office tasks under the setup \cite{hashizume2007multi, cheng2015using}. We aimed to examine if the known EEG signatures of memory workload from literature \cite{lim2018stew, onton2005frontal, brouwer2012estimating, shou2013frontal,puma2018using} are indeed less phenomenal under the dual-monitor setup, which can be physiological evidence of the ergonomic design. 
Another goal of this study is to find more EEG features other than the known signatures from the literature by comparing EEG spectral power, mutual information, and coherence under two situations and using these features to train machine learning models to classify high and low memory workload states.
\color{black}

This study also analyzed EEG data from a Sternberg task recorded in \cite{onton2005frontal} to validate whether the new workload-related EEG signatures found in the study were consistent with those found in the previous study. 

\section{Method}

\subsection{Number-copying Experiment Design}

\begin{figure}[t]
    \centering
    \includegraphics[width=0.6\textwidth]{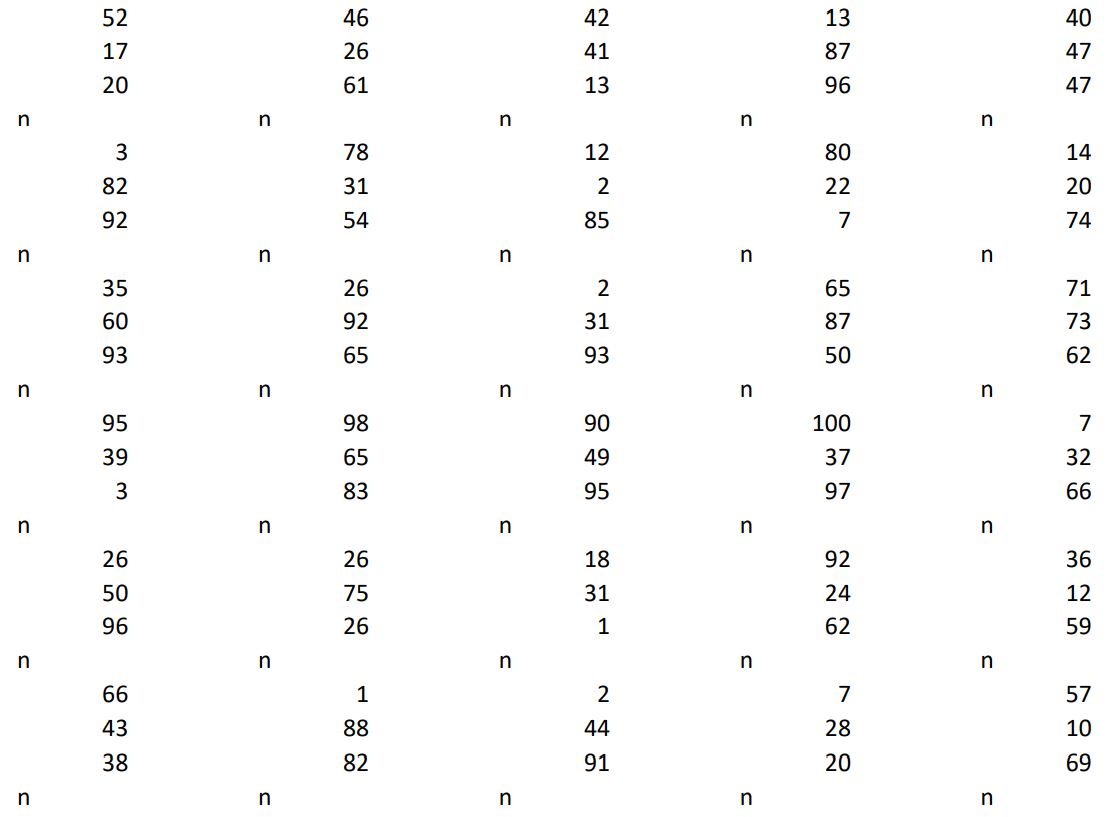}
    \caption{The first page of an example PDF file. The letter n indicated the end of a group of numbers. There were 48 groups of numbers in one file.}
    \label{fig:pdf_example}
\end{figure}

To study the workload while a user is conducting some office work, we designed an experiment in which participants performed a number-copying task under two conditions: single-monitor or dual-monitor configurations. This task seeks to mimic routine office tasks such as accounting and meeting scheduling, which need workers to "copy" and paste numbers between multiple windows or files. Participants were asked to sit in an office chair at a desk and operate a PC with two monitors during the experiment. One monitor (the main/front monitor) was positioned directly in front of the subject and the other one (the side monitor) was positioned to the right of the front monitor and pivoted 30 degrees to face the subject. On the PC, there were numerous PDF files containing a collection of random numbers. Fig. \ref{fig:pdf_example} shows the first page of an example PDF file. The participants had to use the keyboard to enter the same figures into an empty Excel page. Copying and pasting were not allowed. There were two sorts of PDF files in the single-monitor setup: 3 numbers per group (as in the example in Fig. \ref{fig:pdf_example}) and 4 numbers per group. In the single-monitor condition, the side monitor was turned off, and subjects had to memorize several numbers at a time from the PDF file, switch between a PDF file and an excel sheet using Alter+Tab, and enter the numbers into the Excel sheet. In the dual-monitor arrangement, both monitors were turned on and the Excel sheet window was on the front, and the PDF file was on the side monitor. There was no need for subjects to memorize the numbers because they could enter them while looking at a side monitor.  The third type of PDF file was also given to participants in the dual-monitor setup, with half of the groups consisting of three numbers and the other half consisting of four. As a result, there were three different types of PDF files: (1) three numbers per group for the single-monitor setup, (2) four numbers per group for the single-monitor setup, and (3) a combination of three and four numbers per group for the dual-monitor setup.
\color{black}
In the experiment, the subjects had to complete each type of file once as one cycle, and finish two cycles in total. In other words, each subject completed six files (two files per type) in total. Within each cycle, the order was either (1) $\rightarrow$ (3) $\rightarrow$ (2) or (2) $\rightarrow$ (3) $\rightarrow$ (1). That is, a file for the dual-monitor setup was always completed between two files of the single-monitor setup within a cycle. Each file took around 10 minutes to complete.
\color{black}

A total of ten people (4 females and 6 males, aging from 25-30 years old) took part in this investigation. Before participating in the experiment, all individuals were required to read and sign an informed consent form, which had been approved by the Human Research Protections Program at the University of California San Diego under project No. 140053. The SMARTING mobile EEG amplifier (mBrainTrain, Belgrade, Serbia) was used to record EEG data using a Saline-based 24-channel EEG cap. The amplifier's output was transmitted over Bluetooth to the SMARTING acquisition program, and the signals were then broadcasted and recorded using Labstreaminglayer (LSL). Java software collected all of the keyboard inputs and synchronized them with the EEG signals using LSL.

\subsection{Data Processing}
\label{subsection:processing}
All EEG channels were re-referenced to Cz after the EEG signals were filtered with a 0.1-50 Hz band-pass filter. To eliminate EEG artifacts, the function \texttt{clean\_asr} in EEGLAB \cite{delorme2004eeglab} was used to perform artifact subspace reconstruction (ASR) \cite{chang2018evaluation}. The calibration data for ASR were taken from a 30-second eye-close resting interval recorded at the beginning of the experiment. The signals were then downsampled to a sampling rate of 250 Hz.

This study focused on the EEG window preceding the participants' entry of the numbers. During this time, we believe the subjects went through a memory retrieval process to recall the numbers they had just memorized. 
An EEG trial began one second before the first digit of a series of numbers was entered. That is, if a subject's timestamp for entering the first digit of a series of numbers was $t$ second, the EEG trial corresponding to that group was epoched in the window [$t-1$, $t$] sec. As a result, each subject should ideally have 288 trials (6 pdf files $\times$ 48 groups per file). However, several trials were missing from some subjects' experiments due to connection issues during recording. Table \ref{table:subject_trial_nums} lists the final number of trials for each subject. After the EEG trials were epoched, they were concatenated along the time dimension as a long trial. The EEG data were subsequently decomposed into statistically independent components using Independent Component Analysis \cite{makeig1996independent}. We then used the ICLabel package \cite{pion2019iclabel} to identify the artifact and brain components in this long trial. The artifact components were removed, and the brain components were back-projected to the original scalp channels and reshaped back to the same sets of epochs. 

\begin{table}[h!]
\centering
\begin{tabularx}{0.8\textwidth}{ m{5em}  m{8em} | m{5em}  m{8em}}
Subject ID & Number of trials & Subject ID & Number of trials \\ \hline
 1 & 240 & 6 & 288 \\
 2 & 288 & 7 & 96 \\
 3 & 96 & 8 & 192 \\
 4 & 288 & 9 & 282 \\
 5 & 288 & 10 & 282 
\end{tabularx}
\caption{Number of trials in each subject's recording of the number-copying data set.}
\label{table:subject_trial_nums}
\end{table}

\subsection{Sternberg Task Data Set}
\label{subsection:sternberg}
We also used a data set collected during a Sternberg task from a prior study \cite{onton2005frontal} to validate the signatures we discovered in the number-copying data set.
\color{black}
In the Sternberg task, subjects were asked to memorize a sequence of letters. At the beginning of each trial, there was a 5-second eye-open resting period. Then the first letter appeared on a computer screen for 1.4 seconds and disappeared. Right after the first letter disappeared, the second letter appeared for another 1.4 seconds, and so on and so forth until eight letters were presented to the subjects for each trial. After the eighth letter disappeared, the screen stayed blank for a short random time between two to four seconds during which the subjects had to maintain the memory of the sequence of letters they just saw. Finally, after the random blank time, a probe letter was presented, and the subjects were asked to answer whether the probe letter was on the list of letters to memorize in the first part by clicking yes/no buttons.

Although eight letters were presented to the subjects, they didn't need to memorize all of them. Among the eight letters, some of the letters were labeled with green colors, indicating that they should be ignored. The others labeled with black were the ones they actually needed to memorize. The number of the green letters could be six, four, and two, making the number of black letters two, four, and six. We defined the trials with two/four/six black letters as easy/medium/hard trials because with more letters to be memorized, the trials should be more difficult. 
Each subject completed 50-100 trials. Each difficulty accounted for one-third of the trials and was randomly ordered.
Due to missing channel labels in the recordings from some subjects, only eleven subjects' data were included in this study. 
\color{black}

\color{black}
As mentioned in section \ref{subsection:processing}, we focus on the memory retrieval period in this study. To validate the EEG signatures in the comparable two states: performing memory-retrieval (i.e., under single-monitor setup), and not performing memory-retrieval (i.e., under dual-monitor setup),  we made two comparisons of different periods in the Sternberg task.
The first comparison is comparing the period of resting states (not performing memory-retrieval) and the period between the subjects saw a probe and responded (performing memory-retrieval). For each trial, we extracted the eye-open resting period (five-second long) as epochs marked 'Rest'.  As for the recalling period, one-second epochs after the onset of showing the probe letters were extracted and marked 'Recall'.
For the second comparison, we further divided the 'Recall' trials into two labels by the difficulty of the trials. As mentioned above, trials had three levels of difficulty (easy/medium/hard) based on the number of letters to be memorized. We compared the EEG responses of the recalling periods of the easy trials (low memory-retrieval workload) versus the hard trials (high memory-retrieval workload).
\color{black}

\subsection{EEG Spectral Analysis}
Several studies have reported that the EEG power in the delta (1-4 Hz), theta (4-8 Hz), and alpha (8-13 Hz) bands in the frontal area highly correlated with changes in mental and memory workload \cite{lim2018stew, onton2005frontal, brouwer2012estimating, shou2013frontal}. Because some subjects had inconsistent missing channels in both datasets, and we want to preserve as many subjects as possible, only a subset of channels F3, F4, Fz, Pz, AFz, CPz, and POz were analyzed below.  The power of delta, theta, and alpha bands of this subset of channels was extracted by summing the square of the Fourier transform magnitude within corresponding frequency ranges and taking the logarithm of the summation. Totally, 21 features (3 bands $\times$ 7 channels) were extracted from the EEG  spectral analysis and referred to as band power (BP) features in the following sections.

\subsection{Mutual Information}
Mutual Information (MI) is the measure of how two random variables depend mutually. The mutual information of random variables $X$ and $Y$ is defined as:
\begin{eqnarray*}
    & MI(X, Y) \\= & \sum_{y \in \mathcal{Y}} \sum_{x \in \mathcal{X}} p(X=x, Y=y) \log{\frac{p(X=x, Y=y)}{p(X=x)p(Y=y)}},
\end{eqnarray*}
where $\mathcal{X}, \mathcal{Y}$ are the spaces of $X, Y$, $p(X=x,Y=y)$ is the joint probability mass function of $X$ and $Y$, and $p(X=x), p(Y=y)$ are marginal probabilities. In practice, the joint probability and marginal probability are estimated by creating the histograms of observed samples. In EEG analysis, measurements of two channels over a certain period can be used to estimate the probabilities and the mutual information between them.

A variety of EEG paradigms have used mutual information-based features \cite{wang2015using, lan2006salient, jeong2001mutual, blanco2016quantifying}. This study also compared the mutual information between pairs of the same set of selected channels in the spectral analysis under the two monitor setups. However, instead of measuring the mutual information of the raw signals for each channel pair, we used the square of the signals to represent their  power. The number of bins for constructing the histograms was set to 64. An open-source implementation in MATLAB \cite{mijose} was used. 

The calculation of the mutual information value is slightly different between the number-copying data set and the Sternberg data set. In the number-copying data set, the entire 1-sec epochs were used as inputs to the MI calculation. In the Sternberg data set, as the lengths of epochs from the resting and recalling periods were not the same, to avoid the effect of the length differences, we calculated several values of mutual information across an epoch using a one-second non-overlapping sliding window. Then, the final mutual information was obtained by averaging across all five windows.

 Because the symmetrical terms of the mutual information had the same values (e.g., MI of Fz and Pz is equal to MI of Pz and Fz), we only need to calculate 28 features ($7\ {\rm channels} \times (7+1) / 2$), which were referred to as mutual information (MI) features in the following sections.

\subsection{Magnitude-squared Coherence}
Magnitude-squared coherence (coherence) is a metric used to assess the relationship between two signals in the frequency domain. The coherence estimates the quotients between the cross-spectral densities of two signals and the product of their spectral densities. If two time-series signals are defined as $x(t)$ and $y(t)$, the coherence between these two signals $C_{x, y}$ is defined as:
\begin{equation*}
    C_{xy}(f) = \frac{|P_{xy}(f)|^2}{P_{xx}(f)P_{yy}(f)},
\end{equation*}
where $P_{xy}(f)$ is the cross power spectral densities between $x$ and $y$, and $P_{xx}(f)$, $P_{yy}(f)$ are their spectral densities \cite{mandel1976spectral}.

This study first calculated the coherence between each pair of the selected channels (the same pairs as in the spectral and MI analysis) as a function of frequency. Then, we extracted the delta (1-4 Hz), theta (4-8 Hz), and alpha (8-13 Hz) coherence by summing the coherence within the corresponding frequency. Again, to circumvent the effect of length differences, the calculation of the coherence in the Sternberg data set also used the sliding window approach, while the entire epochs were used in the number-copying data set.
It's also worth noting that a channel's coherence with itself is one at all frequencies, so they weren't included in the feature space. In total, 63 features were extracted ($3\ {\rm bands} \times 7\  {\rm channels} \times (7-1) / 2$) and referred to as coherence (Coh) features in the following sections.

\subsection{Evaluation of Classification}
\color{black}
This study aimed to evaluate how well the band power, mutual information, and coherence can track the changes in memory workload levels. We used each of the aforementioned features and a combined feature space to train machine learning models and examined their ability to estimate the participants' memory workload states. As mentioned in section \ref{subsection:processing} and \ref{subsection:sternberg}, we tried to distinguish low/no memory-retrieval workload versus high memory-retrieval workload, so the machine learning models were used to solve the two-class classification problem. Because the design of the classifiers was not the focus of this study, we chose a common type of machine learning model, support vector machines (SVM) as our classifier. We trained the SVM classifiers with four different feature spaces: 1) Band powers only, 2) Mutual information only, 3) Coherence only, and 4) The concatenation of the first three.

For the number-copying data set, in order to avoid temporal auto-correlations between consecutive EEG epochs, a cross-block evaluation within each subject was exploited. That is, at each iteration, epochs of one block (one PDF file) were used as testing data and all the epochs of other blocks were used as training data. No training and testing epochs were extracted within the same block. 
\color{black}
When assessing a subject's data that had no missing values, each iteration involved 240 training epochs and 48 testing epochs
\color{black}

Due to the short duration (one second) of each epoch, the single-epoch classification performance is limited. We used a sample-grouping method to boost the classification performance. Within each iteration of the cross-block validation, after the data were split into training sets and testing sets, every $N_g \in \mathbb{N}$ samples of the same class were averaged and became a new sample. For example, if there are $n$ training samples of one class $j$, denoting as $\mathbf{x}^{(j)}_i \in C_j$, where $i = 1 ... n, C_j \in \{0, 1\}$. The first new training sample after the grouping method will be $\hat{\mathbf{x}}^{(j)}_1 = \sum_{k=1}^{N_g}\mathbf{x}^{(j)}_i/N_g$, and the second one will be $\hat{\mathbf{x}}^{(j)}_2 = \sum_{k=N_g+1}^{2N_g}\mathbf{x}^{(j)}_i/N_g$, and so on and so forth. Note that the order of $i$ followed the same temporal order as they were recorded within each block. Also, if there are remaining samples less than $N_g$ samples, they are averaged and become the last new sample. This sample-grouping method mimics the scenario that the cognitive monitoring system can only make a prediction after measuring a few epochs.  This study showed the results of using different values of $N_g = [1, 2, 4, 8]$.

As for the Sternberg data set, because the epochs of low and high memory workload appeared alternately in time, there should be no effect of temporal auto-correlation, and therefore, simple five-fold cross-validation was employed and repeated twenty times with different random seeds for shuffling.

To ensemble the comparison of sensitivity or specificity, we used the metric balanced accuracy, which is the mean of sensitivity and specificity.  A balanced accuracy averaged across all iterations of cross-validation was obtained for each subject. Finally, the Wilcoxon signed-rank test \cite{woolson2007wilcoxon} was applied to the samples of balanced accuracy of each subject (10 accuracy samples in the number-copying experiment, and 11 samples in the Sternberg task study) to pair-wisely compare the models. Note that the distribution of accuracy is usually non-Gaussian, therefore, the Wilcoxon signed-rank test as a non-parametric paired comparison method was employed.
The Python package Scikit-learn \cite{scikit-learn} was used to implement the machine learning classifiers. Some parameter settings of the classifiers are noteworthy. The SVM models used the Radial Basis Function (RBF) kernel, and the class weights were set to balanced weights (which means the weight of a class is equal to one divided by the frequency of the class). The other parameters of the classifiers simply used the default values. 

\color{black}

\begin{figure}[t]
    \centering
    \includegraphics[width=0.8\textwidth]{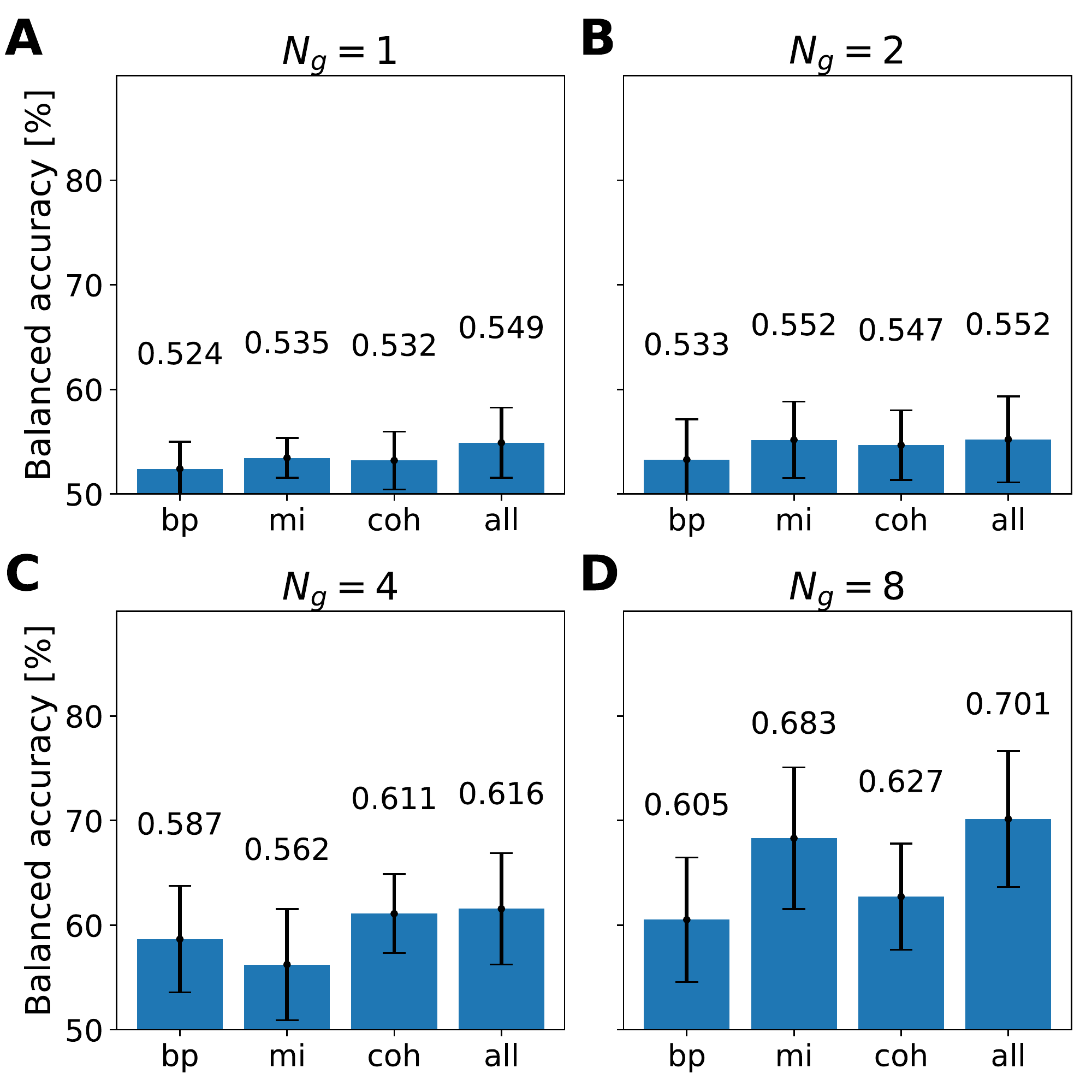}
    \caption{
    \color{black}
    Results of the number-copying experiment. Bar plots of balanced accuracy when using different values of $N_g$. In each panel, the labels of x-axis show different feature spaces: 1) Band power (bp), 2) Mutual information (mi), 3) Coherence (coh), 4) All combined (all). 
    \color{black}
    }
    \label{fig:acc_barplot}
\end{figure}

\begin{figure}[t]
    \centering
    \includegraphics[width=0.5\textwidth]{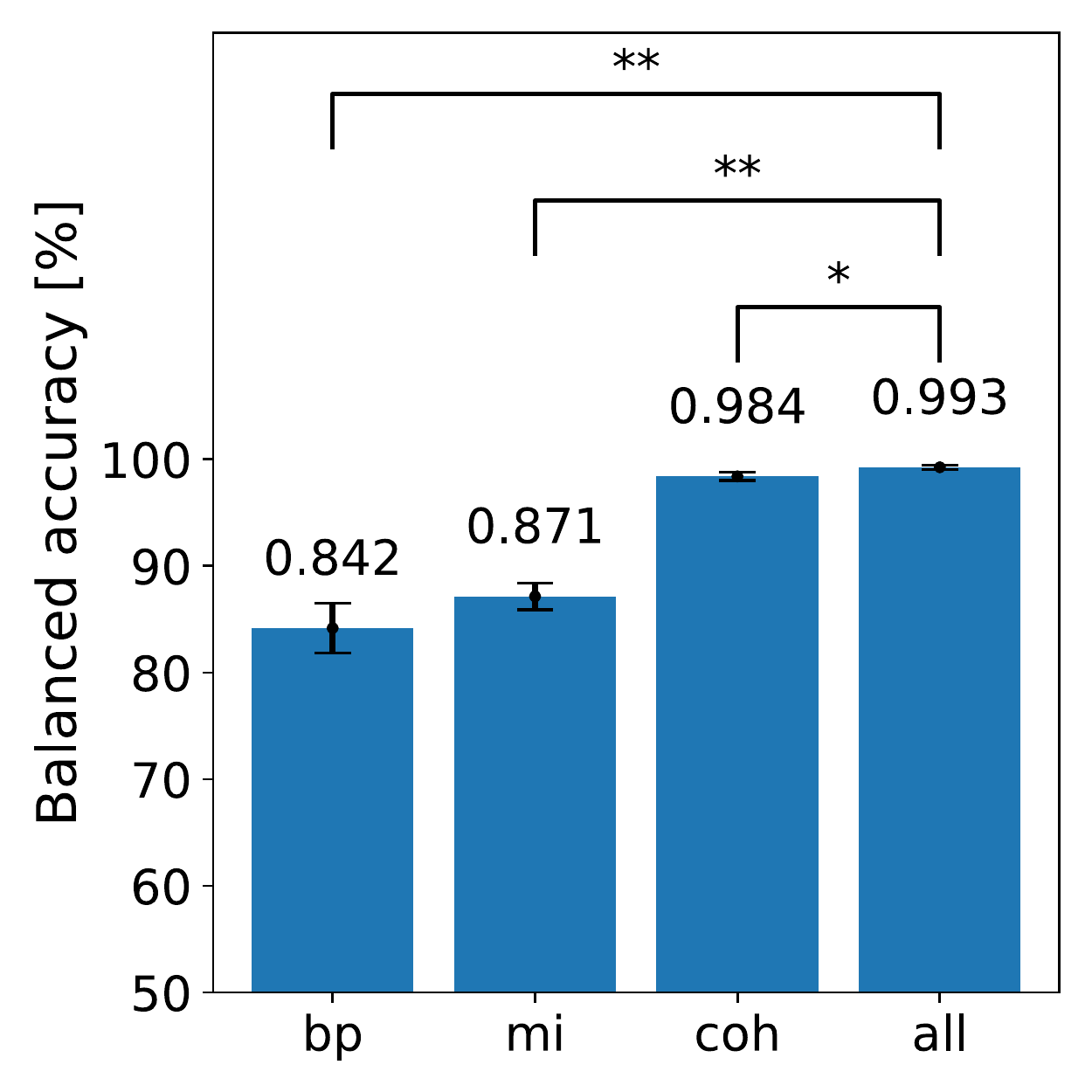}
    \caption{
    \color{black}
    Results of the Sternberg task experiment. Bar plots of balanced accuracy when using different values of $N_g$. In each panel, the labels of x-axis show different feature spaces: 1) Band power (bp), 2) Mutual information (mi), 3) Coherence (coh), 4) All combined (all).
    The p-values of the Wilcoxon signed rank test are shown and the symbol */** indicates the case of a p-value smaller than 0.05/0.01.
    \color{black}
    }
    \label{fig:sternberg_acc_barplot}
\end{figure}

\section{Result}

\begin{figure}[t]
    \centering
    \includegraphics[width=0.6\textwidth]{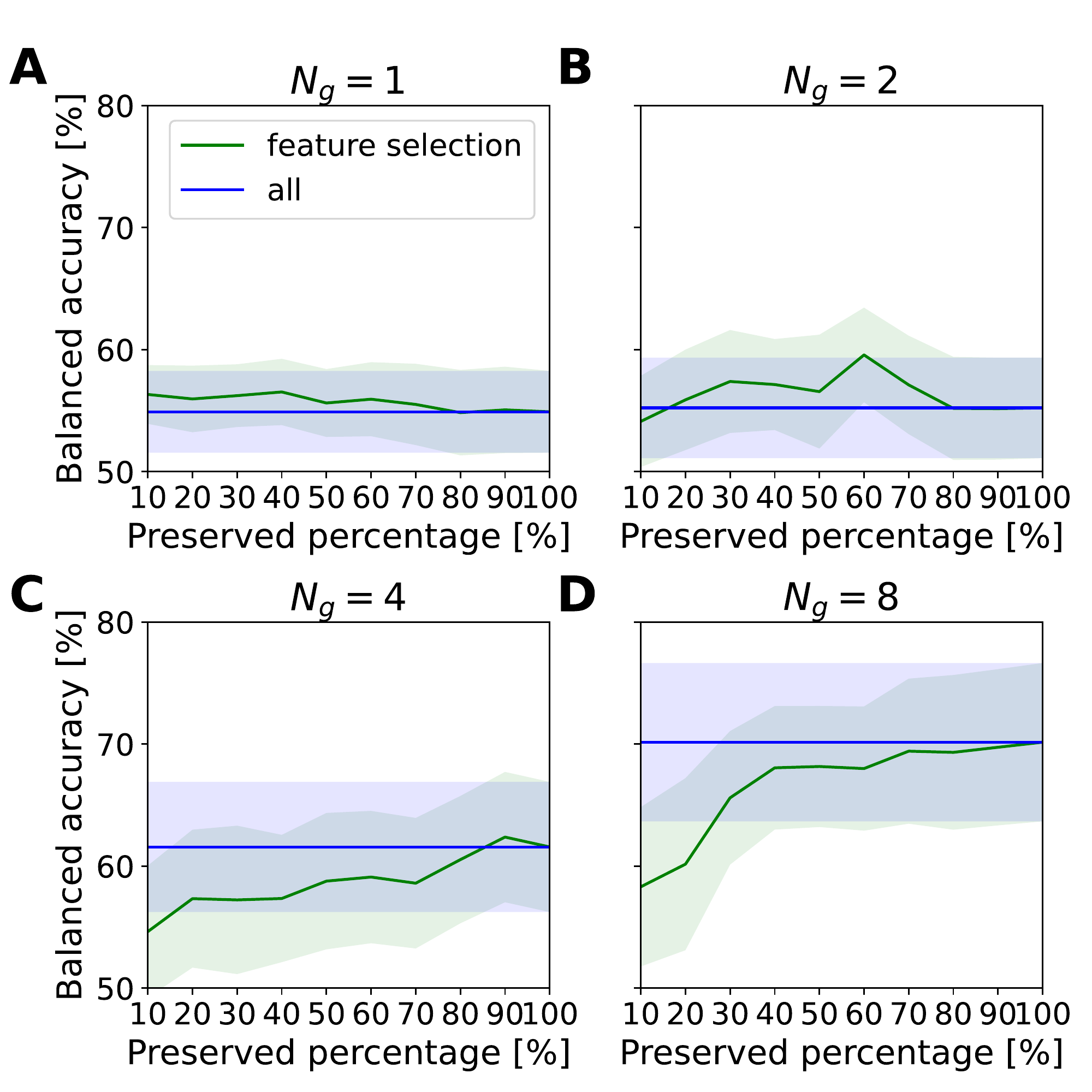}
    \caption{
    \color{black}
    The balanced accuracy versus the percentage of preserved features (from the number-copying data set).
    \color{black}
    }
    \label{fig:sweep}
\end{figure}

\begin{figure}[t]
    \centering
    \includegraphics[width=0.6\textwidth]{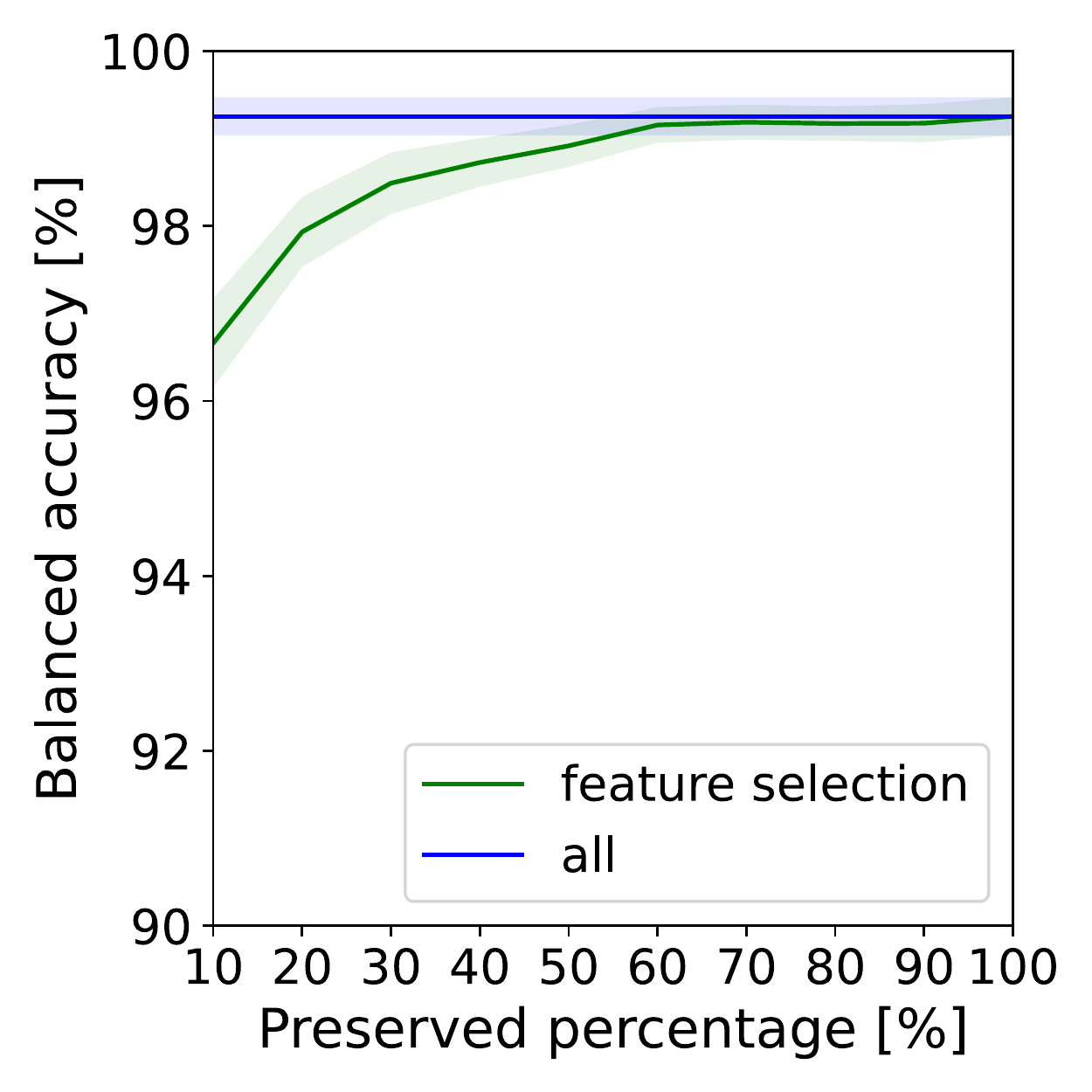}
    \caption{
        \color{black}
        The balanced accuracy versus the percentage of preserved features (from the Sternberg data set).
        \color{black}
        }
    \label{fig:sternberg_sweep}
\end{figure}

\color{black}
Fig. \ref{fig:acc_barplot} shows the results of the classification of the number-copying experiment using different values of $N_g$. Within each panel, the performance of the models trained with four different feature spaces: 1) Band power, 2) Mutual information, 3) Coherence, and 4) All combined is plotted. The heights of the bars represent the average balanced accuracy of 10 subjects. The Wilcoxon signed rank test didn't show a significant difference when comparing models trained with each of the first three feature spaces to those trained with the fourth (all combined). Yet, in all panels, combining all features yields the best results, and in most cases, using only band power features brings the worst. In panels A and B, when $N_g = 1$ and $N_g = 2$, the classification performance is near the chance level (50\%). While in panels C and D, when $N_g = 4$ and $N_g = 8$, the accuracy of using all features combined can go up to 70\%.

Fig. \ref{fig:sternberg_acc_barplot} consists of a similar plot as in Fig. \ref{fig:acc_barplot} using the results from the Sternberg task experiment and classifying the Recall epochs versus Rest. Because the single-epoch classification accuracy is high enough, the sample-grouping method was not employed for this evaluation. Similar to the results of Fig. \ref{fig:acc_barplot}, in Fig. \ref{fig:sternberg_acc_barplot}, using all features combined reaches the best performance, and using only band power features yields the worst. Unlike the results of the number-copying task, the Wilcoxon signed rank test showed a significant difference between using all features combined and the other three feature spaces.

The classification results of classifying Recall epochs with easy difficulty versus hard difficulty are not shown because the performance can only achieve around the chance level. The main reasons are the number of training trials in this scheme is much smaller (only one-third of the case when classifying Recall versus Rest) and the samples of two labels have higher similarity because they are all in recall period despite different levels of difficulty. 

Although combining all types of features brings the best performance, there may be redundant features among each type of feature space. To further investigate this, we exploited a second evaluation scheme in which a subset of features were selected. In this evaluation scheme, the cross-block validation for the number-copying data set and 5-fold validation for the Sternberg task data set was also employed, but within each fold, all features were ranked by ANOVA F-value scores, and then only the top $X$\% features were selected from all features combined. We examined the performance of all four types of classifiers with $X$ ranging from 10 to 100, and the results are plotted in Fig. \ref{fig:sweep} and \ref{fig:sternberg_sweep}.

In Fig. \ref{fig:sweep}, the performance of the classifiers is also limited when $N_g = 1$ and $N_g = 2$, but in panels C and D, with the larger values of $N_g$, we can see a clear trend that the accuracy increased as more features being preserved. Also, in Fig. \ref{fig:sternberg_sweep}, preserving more features brings higher classification accuracy and the green line
reaches its plateau after preserving around 70\% of the features, where the green line in Fig. \ref{fig:sweep}D also reached its plateau around.
\color{black}

\begin{figure}[t]
    \centering
    \includegraphics[width=0.9\textwidth]{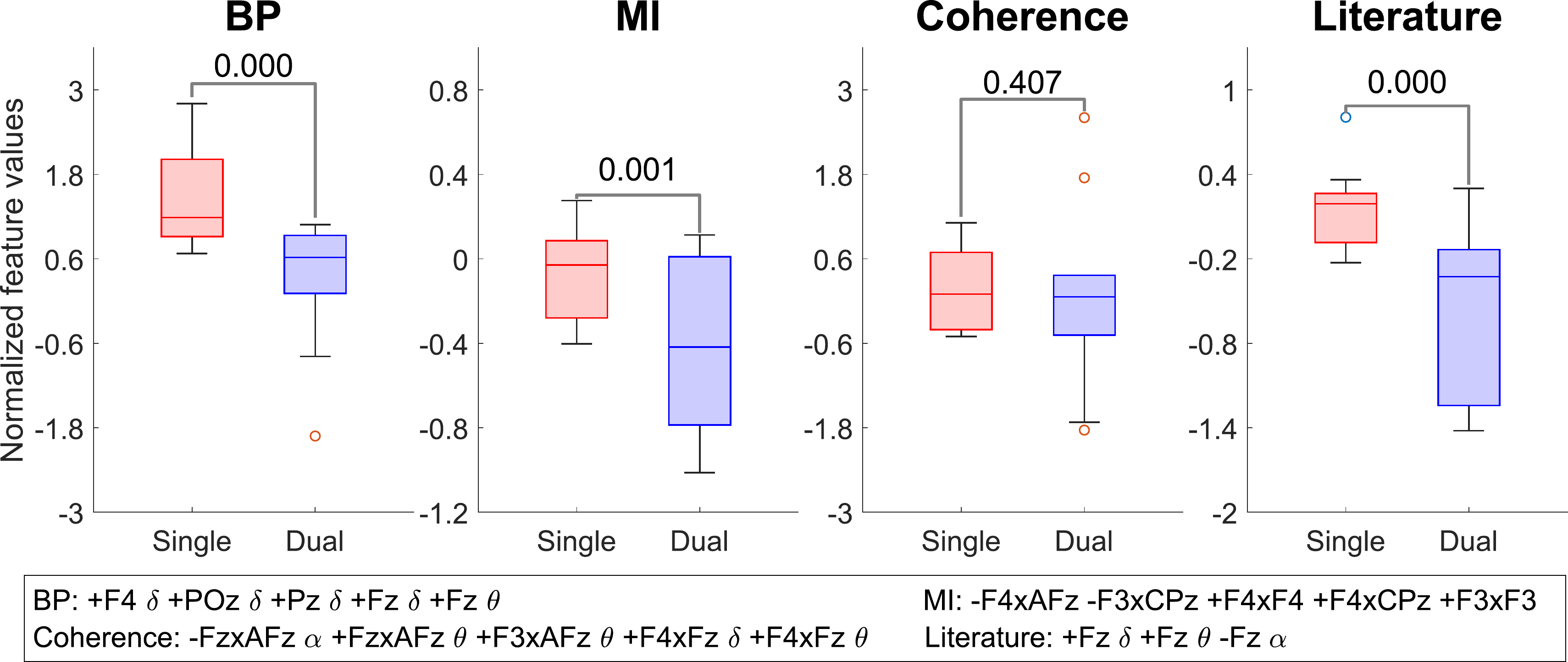}
    \caption{
    \color{black}
    Box plots of the sum of the normalized top feature values across subjects from the number-copying data set. The numbers shown at the top are the p-values of the non-parametric paired bootstrap F-test.
    \color{black}
    }
    \label{fig:single_vs_dual}
\end{figure}

\color{black}
We further examined whether the top features of each feature space selected from the number-copying data set are robust across subjects and across data sets. For each feature space (i.e, BP, MI, and Coherence), we selected the top five features with the highest averaged ANOVA F-value scores (as we used to rank the preserved features in Fig. \ref{fig:sweep} and \ref{fig:sternberg_sweep}) across subjects of the number-copying data set. Within each subject, we normalized each of the top five features to zero mean and unit variance and summed them up for each feature space. Note that when calculating the sum, each normalized feature value could have either positive or negative mean differences between the two classes. Therefore, either $+1$ or $-1$ is multiplied with each normalized feature value to align the polarity. We averaged all summed feature values of the epochs of each label of each subject, and the distribution of the results is shown in Fig. \ref{fig:single_vs_dual}.

\begin{figure}[t]
    \centering
    \includegraphics[width=0.9\textwidth]{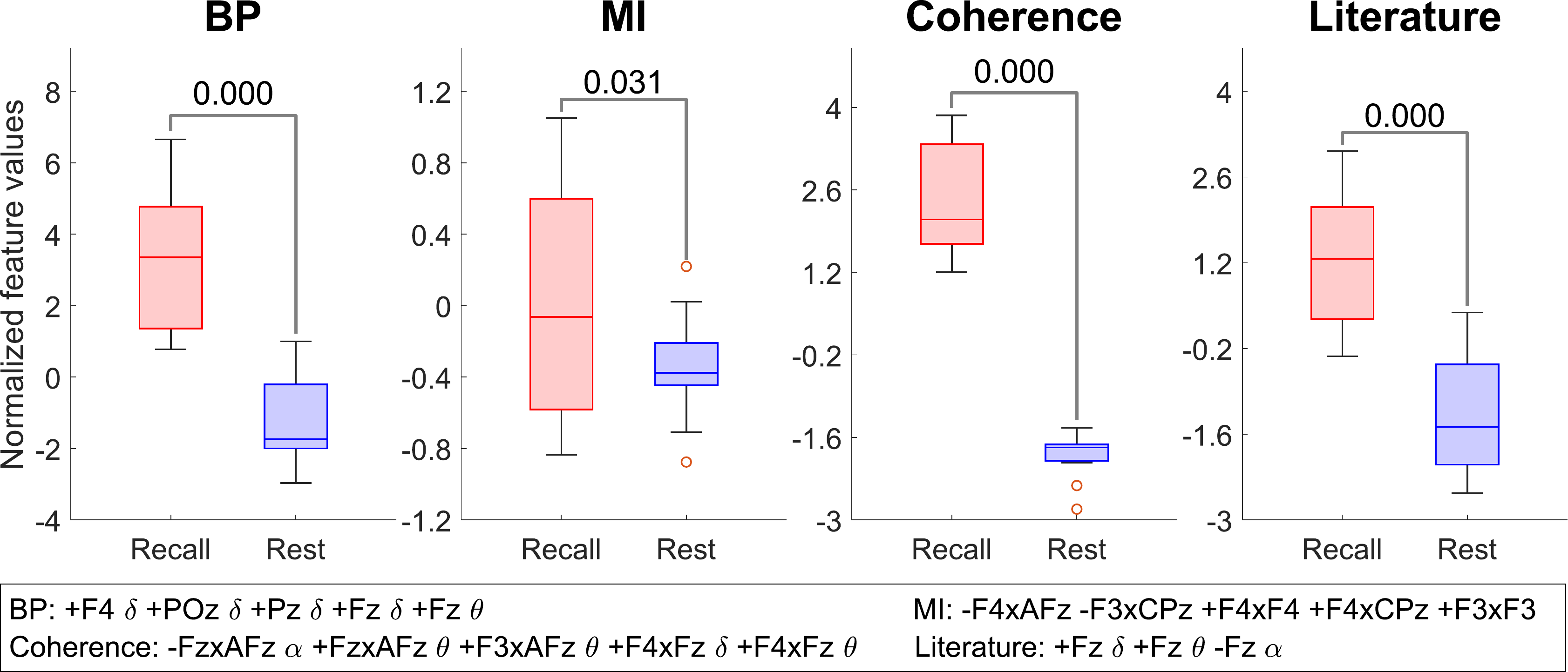}
    \caption{
    \color{black}
    Box plots of the sum of the normalized top feature values across subjects comparing the Recall epochs versus the Rest epochs from the Sternberg task data set.
    \color{black}
    }
    \label{fig:sternberg_recall_vs_rest}
\end{figure}

\begin{figure}[!h]
    \centering
    \includegraphics[width=0.9\textwidth]{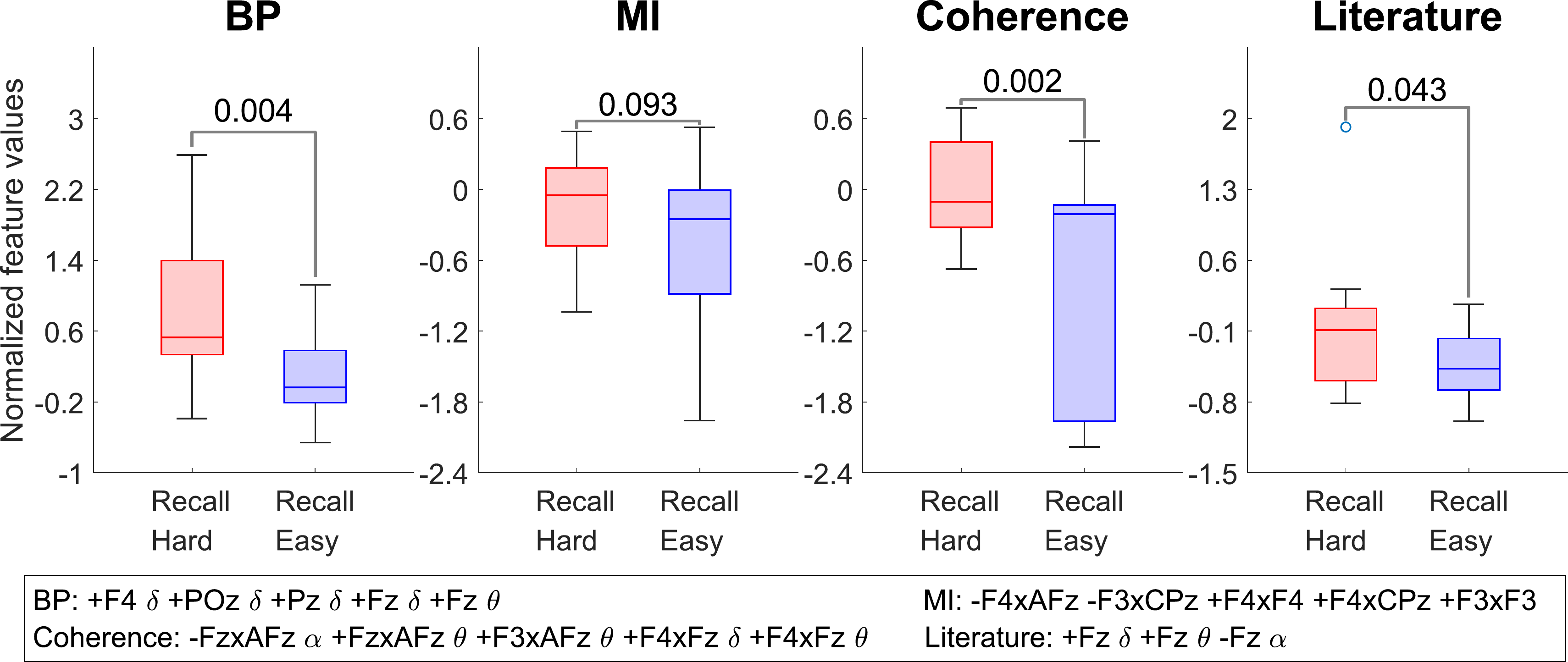}
    \caption{
    \color{black}
     Box plots of the sum of the normalized top feature values across subjects comparing the Recall epochs with the hard difficulty versus the easy difficulty from the Sternberg task data set.
    \color{black}
    }
    \label{fig:sternberg_hard_vs_easy}
\end{figure}

 The box plots in Fig. \ref{fig:single_vs_dual} represent the minimum, first quartile, median, third quartile, maximum values, and outlier values. The text in the lower area shows the selected top features.
Besides the data-driven signatures, we also added a signature based on what previous literature defined as ${\rm Fz}\ \delta+{\rm Fz}\ \theta-{\rm Fz}\ \alpha$ (referred to as the literature signatures) at the right-most column. \cite{lim2018stew, onton2005frontal, brouwer2012estimating, shou2013frontal,puma2018using}.
The differences in the distribution of the two classes were tested with non-parametric paired bootstrap F-test, and the p-values are also shown in Fig. \ref{fig:single_vs_dual}.   The corrected p-values with the Benjamini-Hochberg procedure \cite{benjamini1995controlling} to control the false discovery rate for multiple comparisons are $0.000, 0.002, 0.407$, and $0.000$ for the columns from the left to the right.

Last, similar comparisons for the Sternberg task data set were conducted and the results are plotted in Fig. \ref{fig:sternberg_recall_vs_rest} and \ref{fig:sternberg_hard_vs_easy}. Fig. \ref{fig:sternberg_recall_vs_rest} was plotted to compare the Recall epochs versus the Rest epochs, while Fig. \ref{fig:sternberg_hard_vs_easy} further compared the Recall epochs with the hard difficulty versus the easy difficulty.   Note that these results were obtained using the same sets of normalized selected features with the same polarity as in Fig. \ref{fig:single_vs_dual} in order to validate the data-driven signatures of the number-copying data set.
\color{black}

\section{Discussion}

\color{black}
The results in Fig. \ref{fig:acc_barplot} show that when using $N_g = 8$ the performance for classifying high memory workload state (single-monitor setup) and low memory workload state (dual-monitor setup) can reach around 70\% balanced accuracy. It indicates that in real-world scenarios, the band power, mutual information, and coherence features can be used to detect the changes in memory workload. However, it also shows the limitation that short-epoch prediction (i.e., $N_g = 1$ or $2$) is not plausible using the feature spaces proposed in this study. Only when $N_g = 4$ or $8$, the classification has reasonable performance. Considering each epoch before grouping is one second long, the case $N_g = 8$ is similar to making a prediction with eight-second-long epochs.

We achieved much higher classification accuracy in the Sternberg task data set (Fig. 3) than in our number-coping study; as for the Sternberg task we compared the high memory workload state to a more distinct state, the eye-open resting period. It is important to acknowledge that the Sternberg dataset was collected using a research-grade EEG device in a well-controlled laboratory environment, while our number-copying dataset was collected using a wireless and saline-based EEG headset on freely moving subjects in a real-world setting (a regular office). This difference in signal quality between the datasets may have contributed to the discrepancy in classification accuracy.
\color{black}
Nevertheless, we believe our comparison is still informative, as \cite{wang2019consistency} revealed that the functional connectivity network of the scalp EEG in the fixation state is comparable to that in the working-memory encoding state in an auditory memory span task. While the connectivity patterns are similar, there are quantitative differences between these two states, implying that fixation entails lower cognitive demands (i.e., lower memory workload). Therefore, we believe that distinguishing between fixation and memory retrieval states can be considered akin to distinguishing between low and high memory workload conditions in the number-copying task (Note that in the Sternberg task dataset, we called the eye-open fixation periods "resting" epochs, which corresponds to the "fixation" epochs in the auditory memory span task in \cite{wang2019consistency}.) Regardless of the exact values of the accuracy, the important finding is that the EEG features used in the present study, including band power, mutual information, and coherence, effectively distinguish between high and low memory workload states in both datasets.
\color{black}

Results in Fig. \ref{fig:sweep}, \ref{fig:sternberg_sweep} also imply that most of the features are useful for classifying the states high and low levels of memory workload. In Fig. \ref{fig:sweep}C and D, and \ref{fig:sternberg_sweep}, these models have the best performance when 100\% of features are preserved and preserving less features decreases the accuracy.
Also, in Fig. \ref{fig:sweep}D and \ref{fig:sternberg_sweep}, the accuracy does not increase after preserving more than around 70\%, suggesting that part of the features might be redundant. Yet, including all of the features has no negative impact on performance.

Fig. \ref{fig:single_vs_dual} shows that the summed normalized feature values of all feature space in the single-monitor condition have higher first quartile, median, and third quartile than the dual-monitor condition. Although the statistic test doesn't show a significant difference for the Coherence signatures which can be caused by the effect of the outliers, the differences in the signatures of all other feature spaces are significant. The results suggest that these signatures are robust across subjects.

The literature signatures derived from the frontal activities (the most right panels), which are the most frequently reported features from the literature \cite{lim2018stew, onton2005frontal, brouwer2012estimating, shou2013frontal,puma2018using} significantly increase as the level of memory workload increases. It is proof of the efficacy of our number-copying experiment design.
The selected BP features are also mostly about frontal Delta and Theta power which are similar to the literature signatures with extra terms of parietal Delta. For the MI features, few studies have reported their correlations with memory workload. Mutual information within frontal channels, however, has been reported as a marker of Alzheimer's and schizophrenic disease  \cite{jeong2001mutual, na2002eeg}. The MI signatures we discovered in the current study are also mainly based on the frontal channels. The changes in frontal-parietal mutual information could also result from changes in the frontal and parietal activities reported in  \cite{onton2005frontal, wang2015using}. As for the coherence signatures, they also seem to capture the activities change in the frontal area.

It is not surprising to see significant differences between the feature values of two conditions in Fig. \ref{fig:single_vs_dual} because these are data-driven signatures extracted from this data set. To validate the discovered signatures are not over-fitting features only for the number-copying data set, we tested if the significance still exists in another data set, the Sternberg data set. As shown in Fig. \ref{fig:sternberg_recall_vs_rest} and \ref{fig:sternberg_hard_vs_easy}, these signatures are also robust across subjects in the Sternberg task.
Except for the MI column in Fig. \ref{fig:sternberg_hard_vs_easy}, the statistic tests show significance in all other columns in both figures. For the MI column in Fig. \ref{fig:sternberg_hard_vs_easy}, although the p-value is not below 0.05, it's still quite low (lower than 0.1) considering there are only eleven samples (eleven subjects in the Sternberg data set). The results imply that these signature values robustly increase as the memory workload for each subject increases. Also, we can see that the distributions are more separable in Fig. \ref{fig:sternberg_recall_vs_rest} than in Fig. \ref{fig:sternberg_hard_vs_easy}, this matches the fact that separating performing memory-retrieval (mixed of high and low memory workload) versus resting (zero memory workload) should be easier than separating hard Sternberg tasks (high memory workload) versus easy Sternberg tasks (low memory workload). Although this study has another limitation that the model for single-epoch hard-versus-easy classification couldn't be successfully trained, we still find the significant difference by looking into the average of multiple epochs.

It is noteworthy that even though the non-parametric paired bootstrap F-test indicates a significant difference, some box plots show strongly overlapping distributions between the two conditions. For example, the MI signatures shown in Fig. \ref{fig:single_vs_dual} strongly overlapped in two setups, while the corresponding p-value is less than 0.01. This implies that the distributions of the feature values vary across participants even though the trend is consistent with changes in workload. Therefore, a transfer learning method is required to reduce the variations in the baseline across subjects if one wants to build a plug-and-play workload-monitoring system.

\begin{figure}[t]
    \centering
    \includegraphics[width=0.95\textwidth]{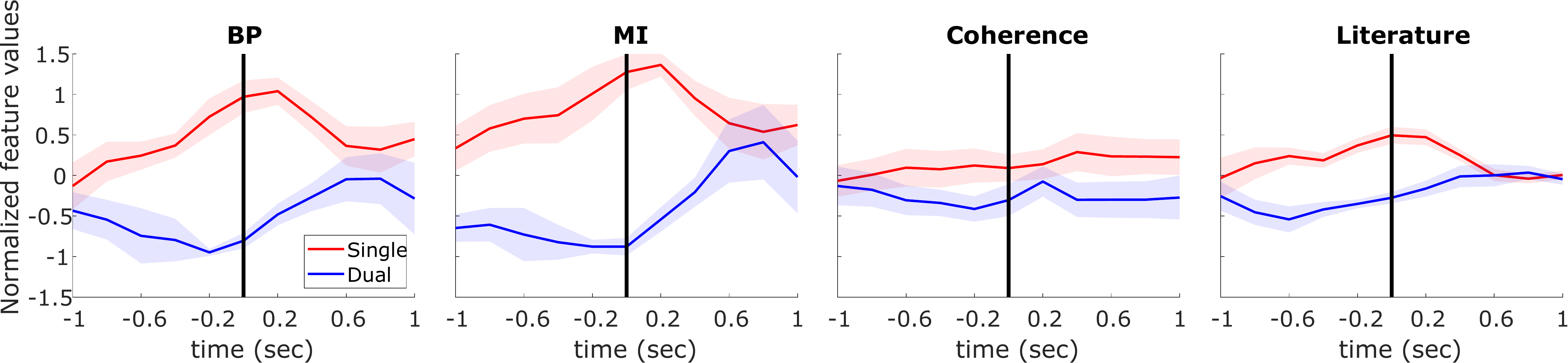}
        \caption{The time-course of the signature values in the number-copying data set. The solid lines are the mean and the shaded areas indicate the standard error across subjects. Trials were aligned to the moment when the first digit of a series of numbers was entered (the black vertical line).}
    \label{fig:dynamic}
\end{figure}

\color{black}
Finally, we checked whether the workload-related signatures we found correlated with the subjects' behavior over time. Fig. \ref{fig:dynamic} shows the dynamics of the BP, MI, Coherence, and the literature-suggested signatures in the number-copying data set. We first aligned all the trials to the moment when the first digit of each group was entered and then calculated the signature values within a 1-sec sliding window stepped at 0.2 seconds (centering at the onset time). Trials were averaged across subjects, and ten subjects' mean and standard errors were plotted. 

\color{black}
Fig. \ref{fig:dynamic} clearly shows that the BP, MI, and literature signatures increase before the onset of entering the first digit under the single-monitor setup and reach their peaks roughly around the onset time. The findings support our hypothesis that the subjects began retrieving numbers before entering them. However, we see a much smaller amplitude of the curve of the coherence signature, indicating that it doesn't have a clear trend. Although the coherence signature didn't correlate with the behavior well, it might be a slowly-reacting response that maintained at higher values during the whole blocks of the single-monitor setup since we also see strong significance in the Sternberg data set.
The signatures in the dual-monitor setup, on the other hand, remain relatively flat before the onset of typing. We also can see an increase in BP, MI, and literature signature values of the dual-monitor setup after the time of tying. Because we don't expect any behavioral changes to coincide with the onset time in this situation, more research is required.

\color{black}
Another limitation of this study is that the size of the number-copying data set is relatively small, and it's mainly because the experiments were conducted during the Covid-19 pandemic.

\section{Conclusion}
This work proposed a new neuroergonomics study that investigates the memory workload in a real-world setting.
\color{black}
Our analysis demonstrated that band power, mutual information, and coherence features are effective in distinguishing between high and low memory workload states.
\color{black}
Furthermore, the data-driven signatures are consistent across both the ecologically valid number-copying task and the standard Sternberg task from a previous study. The findings suggest that EEG recording might be used as an objective measure for human factor investigations in real-world scenarios. 

\section*{Acknowledgment}
The authors would like to express their gratitude to Microsoft Corporation for supporting this work. The authors are also appreciative of all of the volunteers who took part in this study.

This is the Accepted Manuscript version of an article accepted for publication in \textit{Journal of Neural Engineering}. IOP Publishing Ltd is not responsible for any errors or omissions in this version of the manuscript or any version derived from it. The Version of Record is available online at \url{https://doi.org/10.1088/1741-2552/accbed}.

\section*{References}
\bibliographystyle{unsrt.bst}
\bibliography{ref}

\begin{thebibliography}{10}

\bibitem{ms1992human}
EJ~MS and Mccormick Ej.
\newblock Human factors engineering design.
\newblock {\em National Defense Industry Press}, 1992.

\bibitem{mehta2013neuroergonomics}
Ranjana~K Mehta and Raja Parasuraman.
\newblock Neuroergonomics: a review of applications to physical and cognitive
  work.
\newblock {\em Frontiers in human neuroscience}, 7:889, 2013.

\bibitem{wang2015using}
Shouyi Wang, Jacek Gwizdka, and W~Art Chaovalitwongse.
\newblock Using wireless eeg signals to assess memory workload in the $ n
  $-back task.
\newblock {\em IEEE Transactions on Human-Machine Systems}, 46(3):424--435,
  2015.

\bibitem{cegarra2008use}
Julien Cegarra and Aline Chevalier.
\newblock The use of tholos software for combining measures of mental workload:
  Toward theoretical and methodological improvements.
\newblock {\em Behavior Research Methods}, 40(4):988--1000, 2008.

\bibitem{hart1988development}
Sandra~G Hart and Lowell~E Staveland.
\newblock Development of nasa-tlx (task load index): Results of empirical and
  theoretical research.
\newblock In {\em Advances in psychology}, volume~52, pages 139--183. Elsevier,
  1988.

\bibitem{sitaram2007fmri}
Ranganatha Sitaram, Andrea Caria, Ralf Veit, Tilman Gaber, Giuseppina Rota,
  Andrea Kuebler, and Niels Birbaumer.
\newblock Fmri brain-computer interface: a tool for neuroscientific research
  and treatment.
\newblock {\em Computational intelligence and neuroscience}, 2007, 2007.

\bibitem{naseer2015fnirs}
Noman Naseer and Keum-Shik Hong.
\newblock fnirs-based brain-computer interfaces: a review.
\newblock {\em Frontiers in human neuroscience}, 9:3, 2015.

\bibitem{logothetis2008we}
Nikos~K Logothetis.
\newblock What we can do and what we cannot do with fmri.
\newblock {\em Nature}, 453(7197):869--878, 2008.

\bibitem{aghajani2017measuring}
Haleh Aghajani, Marc Garbey, and Ahmet Omurtag.
\newblock Measuring mental workload with eeg+ fnirs.
\newblock {\em Frontiers in human neuroscience}, 11:359, 2017.

\bibitem{bunce2011implementation}
Scott~C Bunce, Kurtulus Izzetoglu, Hasan Ayaz, Patricia Shewokis, Meltem
  Izzetoglu, Kambiz Pourrezaei, and Banu Onaral.
\newblock Implementation of fnirs for monitoring levels of expertise and mental
  workload.
\newblock In {\em International Conference on Foundations of Augmented
  Cognition}, pages 13--22. Springer, 2011.

\bibitem{causse2017mental}
Micka{\"e}l Causse, Zarrin Chua, Vsevolod Peysakhovich, Natalia Del~Campo, and
  Nadine Matton.
\newblock Mental workload and neural efficiency quantified in the prefrontal
  cortex using fnirs.
\newblock {\em Scientific reports}, 7(1):1--15, 2017.

\bibitem{herff2014mental}
Christian Herff, Dominic Heger, Ole Fortmann, Johannes Hennrich, Felix Putze,
  and Tanja Schultz.
\newblock Mental workload during n-back task—quantified in the prefrontal
  cortex using fnirs.
\newblock {\em Frontiers in human neuroscience}, 7:935, 2014.

\bibitem{berka2007eeg}
Chris Berka, Daniel~J Levendowski, Michelle~N Lumicao, Alan Yau, Gene Davis,
  Vladimir~T Zivkovic, Richard~E Olmstead, Patrice~D Tremoulet, and Patrick~L
  Craven.
\newblock Eeg correlates of task engagement and mental workload in vigilance,
  learning, and memory tasks.
\newblock {\em Aviation, space, and environmental medicine}, 78(5):B231--B244,
  2007.

\bibitem{berka2005evaluation}
Chris Berka, Daniel~J Levendowski, Caitlin~K Ramsey, Gene Davis, Michelle~N
  Lumicao, Kay Stanney, Leah Reeves, Susan~Harkness Regli, Patrice~D Tremoulet,
  and Kathleen Stibler.
\newblock Evaluation of an eeg workload model in an aegis simulation
  environment.
\newblock In {\em Biomonitoring for physiological and cognitive performance
  during military operations}, volume 5797, pages 90--99. International Society
  for Optics and Photonics, 2005.

\bibitem{brouwer2012estimating}
Anne-Marie Brouwer, Maarten~A Hogervorst, Jan~BF Van~Erp, Tobias Heffelaar,
  Patrick~H Zimmerman, and Robert Oostenveld.
\newblock Estimating workload using eeg spectral power and erps in the n-back
  task.
\newblock {\em Journal of neural engineering}, 9(4):045008, 2012.

\bibitem{sternberg1966high}
Saul Sternberg.
\newblock High-speed scanning in human memory.
\newblock {\em Science}, 153(3736):652--654, 1966.

\bibitem{bratfisch2008simkap}
O~Bratfisch and E~Hagman.
\newblock Simkap--simultankapazit{\"a}t/multi-tasking.
\newblock {\em M{\"o}dling: Schuhfried GmbH}, 2008.

\bibitem{peck2014using}
Evan~M Peck, Daniel Afergan, Beste~F Yuksel, Francine Lalooses, and Robert~JK
  Jacob.
\newblock Using fnirs to measure mental workload in the real world.
\newblock In {\em Advances in physiological computing}, pages 117--139.
  Springer, 2014.

\bibitem{shou2013frontal}
Guofa Shou and Lei Ding.
\newblock Frontal theta eeg dynamics in a real-world air traffic control task.
\newblock In {\em 2013 35th Annual International Conference of the IEEE
  Engineering in Medicine and Biology Society (EMBC)}, pages 5594--5597. IEEE,
  2013.

\bibitem{di2019eeg}
Gianluca Di~Flumeri, Gianluca Borghini, Pietro Aric{\`o}, Nicolina Sciaraffa,
  Paola Lanzi, Simone Pozzi, Valeria Vignali, Claudio Lantieri, Arianna
  Bichicchi, Andrea Simone, et~al.
\newblock Eeg-based mental workload assessment during real driving: A taxonomic
  tool for neuroergonomics in highly automated environments.
\newblock In {\em Neuroergonomics}, pages 121--126. Elsevier, 2019.

\bibitem{giorgi2021wearable}
Andrea Giorgi, Vincenzo Ronca, Alessia Vozzi, Nicolina Sciaraffa, Antonello
  Di~Florio, Luca Tamborra, Ilaria Simonetti, Pietro Aric{\`o}, Gianluca
  Di~Flumeri, Dario Rossi, et~al.
\newblock Wearable technologies for mental workload, stress, and emotional
  state assessment during working-like tasks: A comparison with laboratory
  technologies.
\newblock {\em Sensors}, 21(7):2332, 2021.

\bibitem{hashizume2007multi}
Ayako Hashizume, Masaaki Kurosu, and Takao Kaneko.
\newblock Multi-window system and the working memory.
\newblock In {\em International Conference on Engineering Psychology and
  Cognitive Ergonomics}, pages 297--305. Springer, 2007.

\bibitem{cheng2015using}
Tsung-Sheng Cheng, Yu-Chun Lu, and Chu-Sing Yang.
\newblock Using the multi-display teaching system to lower cognitive load.
\newblock {\em Journal of Educational Technology \& Society}, 18(4):128--140,
  2015.

\bibitem{lim2018stew}
WL~Lim, O~Sourina, and LP~Wang.
\newblock Stew: simultaneous task eeg workload data set.
\newblock {\em IEEE Transactions on Neural Systems and Rehabilitation
  Engineering}, 26(11):2106--2114, 2018.

\bibitem{onton2005frontal}
Julie Onton, Arnaud Delorme, and Scott Makeig.
\newblock Frontal midline eeg dynamics during working memory.
\newblock {\em Neuroimage}, 27(2):341--356, 2005.

\bibitem{puma2018using}
S{\'e}bastien Puma, Nadine Matton, Pierre-V Paubel, {\'E}ric Raufaste, and
  Radouane El-Yagoubi.
\newblock Using theta and alpha band power to assess cognitive workload in
  multitasking environments.
\newblock {\em International Journal of Psychophysiology}, 123:111--120, 2018.

\bibitem{delorme2004eeglab}
Arnaud Delorme and Scott Makeig.
\newblock Eeglab: an open source toolbox for analysis of single-trial eeg
  dynamics including independent component analysis.
\newblock {\em Journal of neuroscience methods}, 134(1):9--21, 2004.

\bibitem{chang2018evaluation}
Chi-Yuan Chang, Sheng-Hsiou Hsu, Luca Pion-Tonachini, and Tzyy-Ping Jung.
\newblock Evaluation of artifact subspace reconstruction for automatic eeg
  artifact removal.
\newblock In {\em 2018 40th Annual International Conference of the IEEE
  Engineering in Medicine and Biology Society (EMBC)}, pages 1242--1245. IEEE,
  2018.

\bibitem{makeig1996independent}
Scott Makeig, Tzyy-Ping Jung, Dara Ghahremani, and Terrence~J Sejnowski.
\newblock Independent component analysis of simulated erp data.
\newblock {\em Institute for Neural Computation, University of California:
  technical report INC-9606}, 1996.

\bibitem{pion2019iclabel}
Luca Pion-Tonachini, Ken Kreutz-Delgado, and Scott Makeig.
\newblock Iclabel: An automated electroencephalographic independent component
  classifier, dataset, and website.
\newblock {\em NeuroImage}, 198:181--197, 2019.

\bibitem{lan2006salient}
Tian Lan, Deniz Erdogmus, Andre Adami, Misha Pavel, and Santosh Mathan.
\newblock Salient eeg channel selection in brain computer interfaces by mutual
  information maximization.
\newblock In {\em 2005 IEEE Engineering in Medicine and Biology 27th Annual
  Conference}, pages 7064--7067. IEEE, 2006.

\bibitem{jeong2001mutual}
Jaeseung Jeong, John~C Gore, and Bradley~S Peterson.
\newblock Mutual information analysis of the eeg in patients with alzheimer's
  disease.
\newblock {\em Clinical neurophysiology}, 112(5):827--835, 2001.

\bibitem{blanco2016quantifying}
Justin~A Blanco, Michael~K Johnson, Kyle~J Jaquess, Hyuk Oh, Li-Chuan Lo,
  Rodolphe~J Gentili, and Bradley~D Hatfield.
\newblock Quantifying cognitive workload in simulated flight using passive, dry
  eeg measurements.
\newblock {\em IEEE Transactions on Cognitive and Developmental Systems},
  10(2):373--383, 2016.

\bibitem{mijose}
Jose Delpiano.
\newblock Fast mutual information of two images or signals.
\newblock {M}ATLAB Central File Exchange. Retrieved August 20 2021.

\bibitem{mandel1976spectral}
L~Mandel and E~Wolf.
\newblock Spectral coherence and the concept of cross-spectral purity.
\newblock {\em JOSA}, 66(6):529--535, 1976.

\bibitem{woolson2007wilcoxon}
RF~Woolson.
\newblock Wilcoxon signed-rank test.
\newblock {\em Wiley encyclopedia of clinical trials}, pages 1--3, 2007.

\bibitem{scikit-learn}
F.~Pedregosa, G.~Varoquaux, A.~Gramfort, V.~Michel, B.~Thirion, O.~Grisel,
  M.~Blondel, P.~Prettenhofer, R.~Weiss, V.~Dubourg, J.~Vanderplas, A.~Passos,
  D.~Cournapeau, M.~Brucher, M.~Perrot, and E.~Duchesnay.
\newblock Scikit-learn: Machine learning in {P}ython.
\newblock {\em Journal of Machine Learning Research}, 12:2825--2830, 2011.

\bibitem{benjamini1995controlling}
Yoav Benjamini and Yosef Hochberg.
\newblock Controlling the false discovery rate: a practical and powerful
  approach to multiple testing.
\newblock {\em Journal of the Royal statistical society: series B
  (Methodological)}, 57(1):289--300, 1995.

\bibitem{wang2019consistency}
Ruimin Wang, Sheng Ge, Noha~Mohsen Zommara, Karine Ravienna, Teodora Espinoza,
  and Keiji Iramina.
\newblock Consistency and dynamical changes of directional information flow in
  different brain states: A comparison of working memory and resting-state
  using eeg.
\newblock {\em NeuroImage}, 203:116188, 2019.

\bibitem{na2002eeg}
Sun~Hee Na, Seung-Hyun Jin, Soo~Yong Kim, and Byung-Joo Ham.
\newblock Eeg in schizophrenic patients: mutual information analysis.
\newblock {\em Clinical Neurophysiology}, 113(12):1954--1960, 2002.

\end{thebibliography}

\end{document}